\documentclass[journal,titlepage]{IEEEtran}
\usepackage{ifpdf}
 \ifpdf

 \else
 \fi

\usepackage{cite}

\ifCLASSINFOpdf
   \usepackage[pdftex]{graphicx}
\else
\fi
\usepackage{amsmath}
\usepackage{algorithmic}

\usepackage{array,multirow,colortbl}

\ifCLASSOPTIONcompsoc
  \usepackage[caption=false,font=normalsize,labelfont=sf,textfont=sf]{subfig}
\else
  \usepackage[caption=false,font=footnotesize]{subfig}
\fi
\begin{document}

\begin{titlepage}
{\bfseries\LARGE IEEE Copyright Notice\par}
\vspace{1cm}
{\scshape\Large © 2021 IEEE.  Personal use of this material is permitted.  Permission from IEEE must be obtained for all other uses, in any current or future media, including reprinting/republishing this material for advertising or promotional purposes, creating new collective works, for resale or redistribution to servers or lists, or reuse of any copyrighted component of this work in other works. \par}
\vspace{1cm}
{\Large This is the author's version of an article that has been published in this journal. Changes were made to this version by the publisher prior to publication. \par}
\vspace{1cm}
{\Large The final version of record is available at http://dx.doi.org/10.1109/TIM.2021.3063772 \par}
\end{titlepage}

%
\title{Triangular phase-shift detector for drone precise vertical landing RF systems}
%
%
%

\author{V\'ictor Ara\~na-Pulido,~\IEEEmembership{Member,~IEEE,}
        Eugenio Jim\'enez-Ygu\'acel,~\IEEEmembership{Member,~IEEE,}
        Francisco Cabrera-Almeida,~\IEEEmembership{Member,~IEEE,}
        Pedro Quintana-Morales
\thanks{The authors are with the Institute for Technological Development and Innovation in Communications (IDeTIC), 
Department of Signals and Communication, University of Las Palmas de Gran Canaria (ULPGC), 35017 Las Palmas, Spain
(e-mail: victor.arana@ulpgc.es; eugenio.jimenez@ulpgc.es; francisco.cabrera@ulpgc.es; pedro.quintana@ulpgc.es).}
\thanks{This work was supported by the Spanish Government under Grant TEC2017-88242-C3-3-R Project.}

\thanks{Manuscript received XX XX, 2020; revised XX XX, 202X.}}

%
%

\markboth{IEEE Transactions on Instrumentation and Measurement,~Vol.~XX, No.~X, August~202X}%
{Shell \MakeLowercase{\textit{et al.}}: Bare Demo of IEEEtran.cls for IEEE Journals}
%



\maketitle

\begin{abstract}
This paper presents a circuit for precise vertical landing of drones based on a three phase-shifts detection of a single frequency transmitted from the landing point. The circuit can be considered as a new navigation sensor that assists in guidance corrections for landing at a specific point. The circuit has three inputs to which the signal transmitted from an oscillator located at the landing point arrives with different delays. The input signals are combined in pairs in each of the three analog phase detectors, after having passed through 3 dB@90\textordmasculine\ hybrid couplers that guarantee a theoretical non-ambiguous phase-shift range of $\pm$90\textordmasculine. Each output has a voltage that is proportional to the phase-shift between each of the input signals, which in turn depend on the position relative to the landing point. A simple landing algorithm based on phase-shift values is proposed, which could be integrated into the same flight control platform, thus avoiding the need to add additional processing components. To demonstrate the feasibility of the proposed design, a triangular phase-shift detector prototype has been implemented using commercial devices. Calibration and measurements at 2.46 GHz show a dynamic range of 30 dB and a non-ambiguous detection range of $\pm$80\textordmasculine\ in the worst cases. Those specs let us to track the drone during the landing maneuver in an inverted cone formed by a surface with a $\pm$4.19 m radius at 10m high and the landing point.
\end{abstract}


%
\IEEEpeerreviewmaketitle

\section{Introduction}
\IEEEPARstart{P}{recise} point landing is required by many drone applications (multi-rotors) with autonomous operation: mobile platforms, recharging systems, small enclosures with automated open/close gates where they are stored and protected from adverse environmental conditions, etc. This maneuver is one of the most complex, even worse when it is done autonomously, since it requires large volumes of data to make corrections quickly and accurately \cite{Gautam2014}. The difficulty increases when it is done outdoors since aspects related to visibility (global/local positioning systems, targets, etc.), type of terrain, air turbulence, etc. must be considered. Outdoor systems are also exposed to interference which makes the problem more difficult, as complex processing is required \cite{Shakhatreh2019}. Solutions based on GPS positioning and image processing are the most used. Systems based on pattern image processing located at landing point could solve the problem in adverse conditions but would require a powerful computer platform to provide sufficient ODR (Output Data Rate) to correct the trajectory \cite{Patruno2019,Pluckter2018}. Especially critical are the landing approach maneuvers when it comes to recharge systems where the drone must be placed on certain connection points \cite{Woo2017} and with a certain alignment \cite{Junaid2017}, even worse when it is the drone that acts to recharge other systems \cite{Basha2015}. These maneuvers usually use systems based on satellite positioning and image processing (targets) that can be combined with mobile platforms that allow the adjustment of the recharge system \cite{Choi2016}, allowing greater error tolerance over the landing point. To avoid mechanical systems and improve inductive charging efficiency, solutions using a double inductive ring \cite{Angrisani2015} have been proposed. 

RF (Radio Frequency) systems that provide positions to vehicles, need to locate reference stations to perform triangulation (GPS, radio beacons, etc.). These solutions can improve resolution from phase information and be used for precision landings \cite{Supej2014,Koo2017,Lim2018a}. Phase-shift has also been used in precision mechanical positioning control \cite{Gao2015}. However, they present problems when the vehicle is operating in a terrain with abrupt orography and especially if there are abundant trees that absorb the signal the higher the frequency \cite{Meng2009,Meng2010,Dyukov2016}.

The autonomous landing requires more data for higher intensities of turbulence in the approach, as well as data from the inertial system in relation to the stabilization of the drone when it is suspended in the air. In fact, inertial systems can have ODR in the order of 100 corrections per second \cite{Lim2018b,Hoflinger2012,Garcia2017}, while image processing corrections are running at best 30 corrections per second but usually not more than 10 corrections per second \cite{Patruno2019,Pluckter2018}. Therefore, a system based on image processing is not the most suitable in these conditions. Even under better conditions, an RTK (Real Time Kinematic) system would not be suitable either since they have a low position data rate \cite{Supej2014,Koo2017}, as with systems based on FMCW (Frequency Modulated Continuous Wave) radar \cite{Dobrev2018}, especially when passive reflectors are used \cite{Doer2020}.

There are other methods that are primarily focused on winged aircraft that also require signal clearing, in addition to requiring auxiliary equipment mounted on the ground that complicates operation. Thus, there are those based on IR stereo \cite{Kong2013,Ma2016} that use two beams of infrared signal emitted from two known locations, or those based on FMCW radar that use several transponders \cite{Pavlenko2019}. Another variant uses a single beam but achieves poorer landing accuracy \cite{Lee2019}. Finally, it is worth mentioning the systems in which the drone is attached to a mobile landing platform and then the cable is spooled \cite{Alarcon2019}.

To address the problem of guidance in the final phase of approach and landing, an RF circuit is proposed that allows a control based on phase-shift detection of three input signals from an oscillator located at the landing point. The proposed solution provides DC voltages proportional to drone position relative to landing point. This circuit can be considered as a new navigation sensor that assists in precision landing and facilitates guidance tasks \cite{Gautam2014}. In addition, a very simple algorithm is detailed to perform the landing maneuver that could be integrated into the same flight control platform, thus avoiding the need to add additional processing units. 

First, a simple analysis is carried out to understand the basis on which the proposal is based. Then, the relationship between the detected phase-shift and guidance instructions during the landing maneuver is established. The following section continues with the design of the triangular phase-shift detector prototype. Finally, calibration and experimental characterization is performed to determine non-ambiguous phase-shift operating ranges, inherent in any phase-dependent solution, as well as validation of the proposed algorithm. 

\section{Triangular Phase-Shift Landing System}
\subsection{Fundamentals}
\begin{figure}[!t]
\centering
\includegraphics[width=7cm]{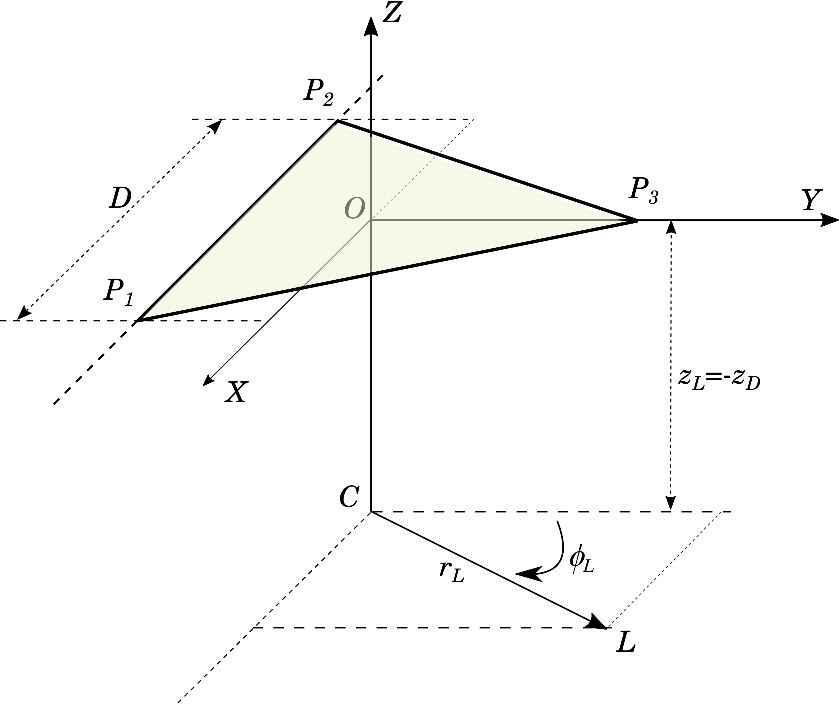}
\caption{Coordinate reference system and points that define the problem.}
\label{fig:figura1}
\end{figure}

Let us assume an environment free of obstacles and multiplath effects. Under these conditions, if a signal of frequency $f$ is transmitted at speed $c$ from a point $L$ towards two points in space ($P_i$ and $P_j$), the time-delay ($\Delta t$) and phase-shift ($\Delta\theta$) of the signals received on $P_i$ and $P_j$ depend on the path difference ($\Delta d_{ij}$)
(Eq \ref{eq:deltaphi}).
\begin{equation}
\label{eq:deltaphi}
\begin{split}
\Delta t_{ij} & = \Delta d_{ij}/c \qquad\quad\ \ [sec]\\
\Delta \theta_{ij} & = 2 f \Delta d_{ij} 180 / c \quad [deg]
\end{split}
\end{equation}
Let us consider a coordinate reference system as shown in Fig. \ref{fig:figura1}, where the drone is located at point $O$ and the landing point is $L$. A signal of frequency $f$ is transmitted from point $L$. The drone receiver has three inputs ($P_1$, $P_2$ and $P_3$) separated by a distance $D$. Point $O$ corresponds to the incenter of the equilateral triangle ($|\overline{OP_1}|$ = $|\overline{OP_2}|$ = $|\overline{OP_3}|$).

The inputs are positioned with point $P_3$ at the drone forward direction. The drone rotation axis corresponds to the $Z$ detected axis: right turn, from  $Y$ axis to $X$ axis and left turn, from $Y$ axis to $-X$ axis. The drone is stabilized in the $XY$ plane and height $z_D$. The landing point ($L$) is located at cylindrical coordinates \{$r_L$, $\phi_L$, $-z_D$\}. 

The coordinates of the entry points ($P_i$) to the circuit according to the unit vectors \{$\hat{x}$, $\hat{y}$,  $\hat{z}$\} are:
\begin{eqnarray}
 \overline{OP_1} & = &\frac{D}{2}\hat{x} - \frac{D}{2}\tan\frac{\pi}{6}\hat{y} \nonumber\\
 \overline{OP_2} & = & \frac{-D}{2}\hat{x} - \frac{D}{2}\tan\frac{\pi}{6}\hat{y}\\
 \overline{OP_3} & = & 0\hat{x} + \frac{D}{2\cos(\pi/6)}\hat{y} \nonumber
 \end{eqnarray}

The coordinate of the landing point is:

\begin{equation}
\overline{OL} = r_L \sin(\phi_L)\hat{x} + r_L \cos(\phi_L)\hat{y} - z_D\hat{z}
\end{equation}

The path differences between two input signals, which are proportional to the phase-shifts (equation \ref{eq:deltaphi}) are:
\begin{eqnarray}
\label{eq:deltas}
 \Delta d_{12} & = & | \overline{OL} - \overline{OP_1} | - | \overline{OL} - \overline{OP_2} | \nonumber\\
 \Delta d_{23} & = & | \overline{OL} - \overline{OP_2} | - | \overline{OL} - \overline{OP_3} |\\
 \Delta d_{31} & = & | \overline{OL} - \overline{OP_3} | - | \overline{OL} - \overline{OP_1} | \nonumber
\end{eqnarray} 
The sum of the path differences is zero ($\sum \Delta d_{ij}=0$) and therefore the sum of the phase-shifts is also zero. According to this, it is enough to know two data points to know the third datum. However, the three data are kept for greater redundancy and providing greater robustness against deviations from ideal behavior of real circuits.

When point $O$ (drone location) is above the perpendicular to point $L$, $|\overline{OP_1}|$ = $|\overline{OP_2}|$ = $|\overline{OP_3}|$ and  $\Delta \theta_{ij}= 0$  are fulfilled.
\subsection{Phase-shifts vs drone location}
\label{subseq:Phsvsdrlo}
\begin{figure}[!t]
\centering
\subfloat[]{\includegraphics[width=2.5cm]{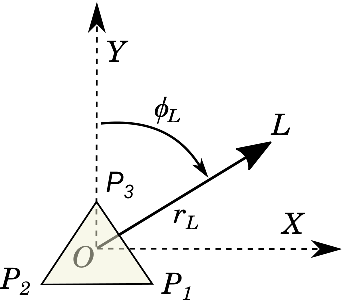}
\label{fig:figura2a}}
\subfloat[]{\includegraphics[width=6.1cm]{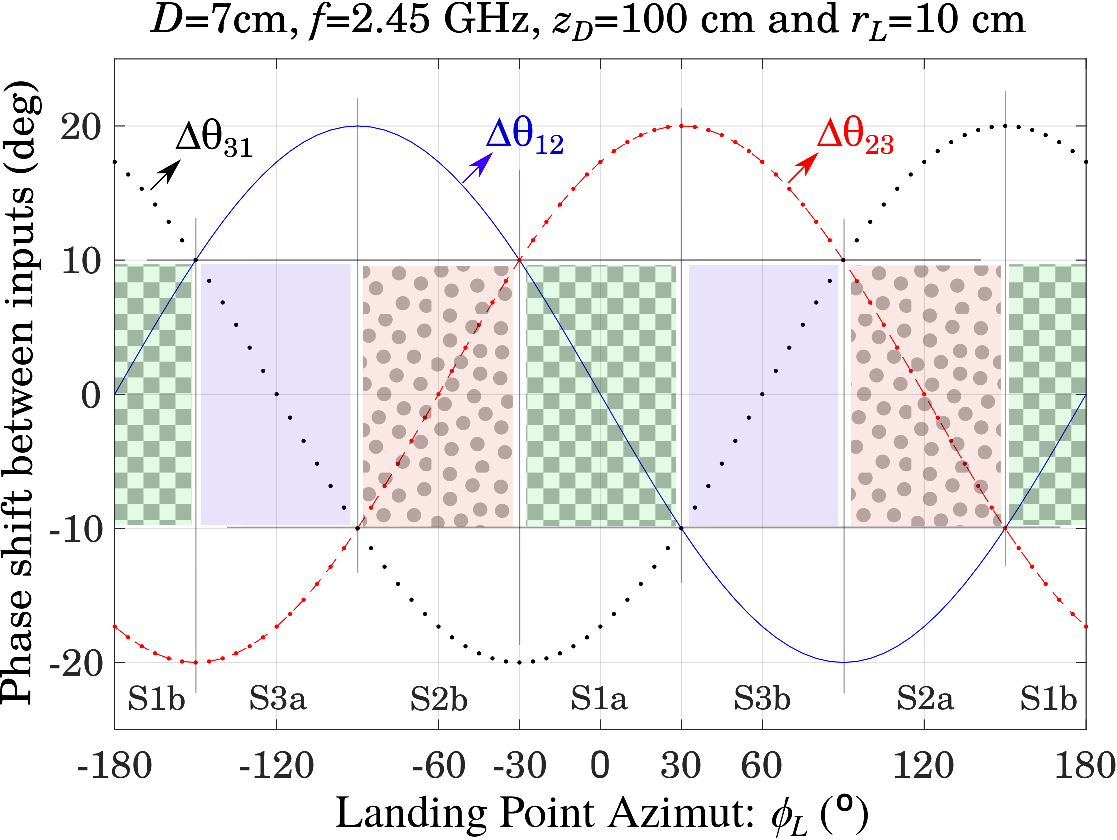}
\label{fig:figura2b}}
\caption{\textbf{(a)} Top view of Fig. \ref{fig:figura1}. \textbf{(b)} Phase shift between input and signals when location of the landing point ($L$) is changed around drone. }
\label{fig:figura2}
\end{figure}

The characterization of phase-shifts as a function of drone location is done by moving the landing point around the drone ($\phi_L$) for a given radius $r_L$. Fig. \ref{fig:figura2a} shows the procedure seen from above (top view from Fig. \ref{fig:figura1}) and phase-shifts between inputs when landing point is rotated from -180\textordmasculine\ to +180\textordmasculine\ are shown in Fig. \ref{fig:figura2b}. 

The results depend on the following parameters where experimental data are between parenthesis: distance between input points ($D$=7cm), signal frequency ($f$=2.45 GHz), drone height ($z_D$=100cm) and distance from drone to landing point ($r_L$=10cm). Attending to the intersection points between curves, the graph can be divided into 6 sectors. In turn, the two sectors of each phase-shift curve ($\Delta \theta_{ij}$) that have values below the crossing point (approximately -10\textordmasculine\ and +10\textordmasculine\ in this case) have been highlighted in Fig. \ref{fig:figura2b}. 

Thus, Sector 1 covers the values between -30\textordmasculine\ and +30\textordmasculine\ (Sector 1a), and between -150\textordmasculine\ and +150\textordmasculine\ (Sector 1b). Sector 1a is in the direction of $\overline{OP_3}$ ($\phi_L$=0\textordmasculine\ in Fig. \ref{fig:figura2a}) and Sector 1b is shifted 180\textordmasculine\ from the previous one. Similarly, Sector 2a is defined between +105\textordmasculine\ and +135\textordmasculine\ (in the direction $\overline{OP_1}$, $\phi_L$= 120\textordmasculine) and Sector 2b is shifted 180\textordmasculine\ (-90\textordmasculine\ to -30\textordmasculine); Sector 3a is defined between -150\textordmasculine\ and -90\textordmasculine\ (in the direction $\overline{OP_2}$, $\phi_L$= \mbox{-120}\textordmasculine) and Sector 3b is shifted 180\textordmasculine\ (+30\textordmasculine\ to +90\textordmasculine). According to this, if the landing point is in Sector 2, rotating the drone +120\textordmasculine\ or -60\textordmasculine\ is enough to locate it in the Sector 1. Likewise, -120\textordmasculine\ or +60\textordmasculine\ when $L$ is in Sector 3.

\subsection{Analog phase-shift detector and non-ambiguous range}
\begin{figure}[!t]
\centering
\subfloat[]{\includegraphics[width=4.2cm]{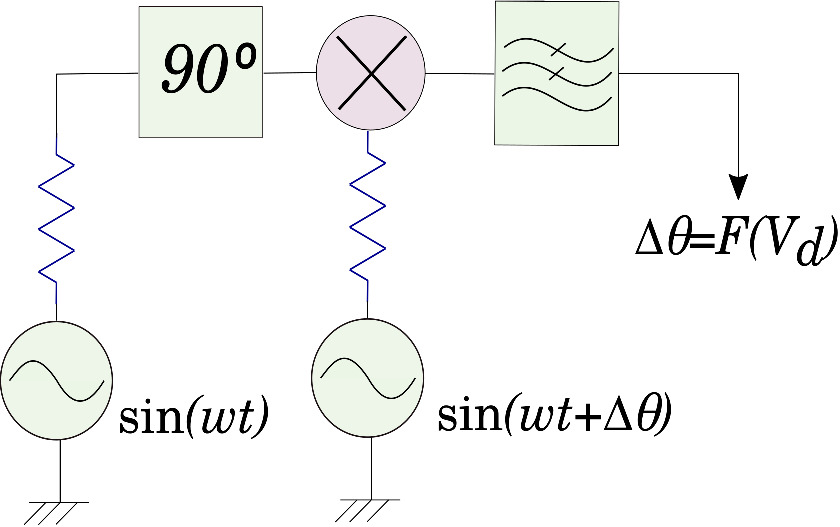}%
\label{fig:figura3a}}
\hfil
\subfloat[]{\includegraphics[width=4.2cm]{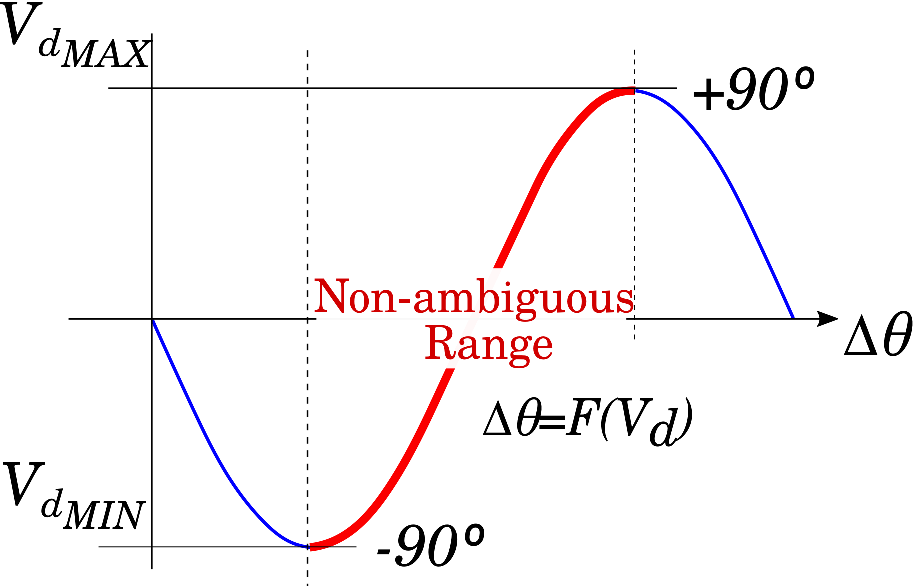}%
\label{fig:figura3b}}
\caption{\textbf{(a)} Simplified measurement system for phase-shift measurements based of analog multiplier. \textbf{(b)} Ideal phase detector response highlighting the non-ambiguous range ($\pm$90\textordmasculine).}
\label{fig:figura3}
\end{figure}

The measurement of the gaps between the three input signals to the circuit is carried out by a phase detector. To measure phase-shifts between $\pm$90\textordmasculine, using 0\textordmasculine\ as a reference, it is common to use a 90\textordmasculine\ phase shifter, a multiplier and a low-pass filter (Fig. \ref{fig:figura3a}). The maximum detectable phase-shift is given by the non-ambiguous zone within the $\pm$90\textordmasculine\ range (Fig. \ref{fig:figura3b}). By assuming $V_{d_{ij}}$=$K_d$ $\sin(\Delta \theta_{ij})$ and solving for $\Delta \theta_{ij}$, the phase-shift can be calculated as: $\Delta \theta_{ij}$ = $F(V_d)$ = $K_d$ $\arcsin (V_{d_{ij}})$, where $|\Delta \theta_{ij}|\leq$ 90\textordmasculine.

\begin{figure}[!t]
\centering
\subfloat[]{\includegraphics[width=4.4cm]{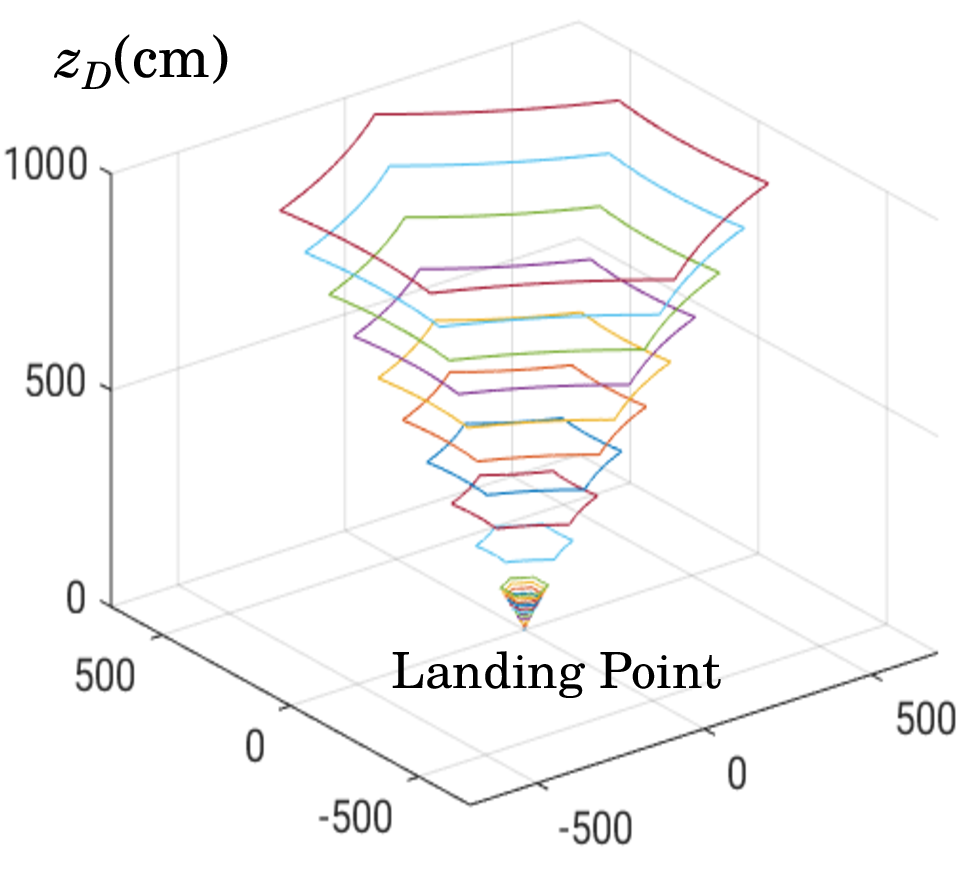}%
\label{fig:figura4a}}
\subfloat[]{\includegraphics[width=4.4cm]{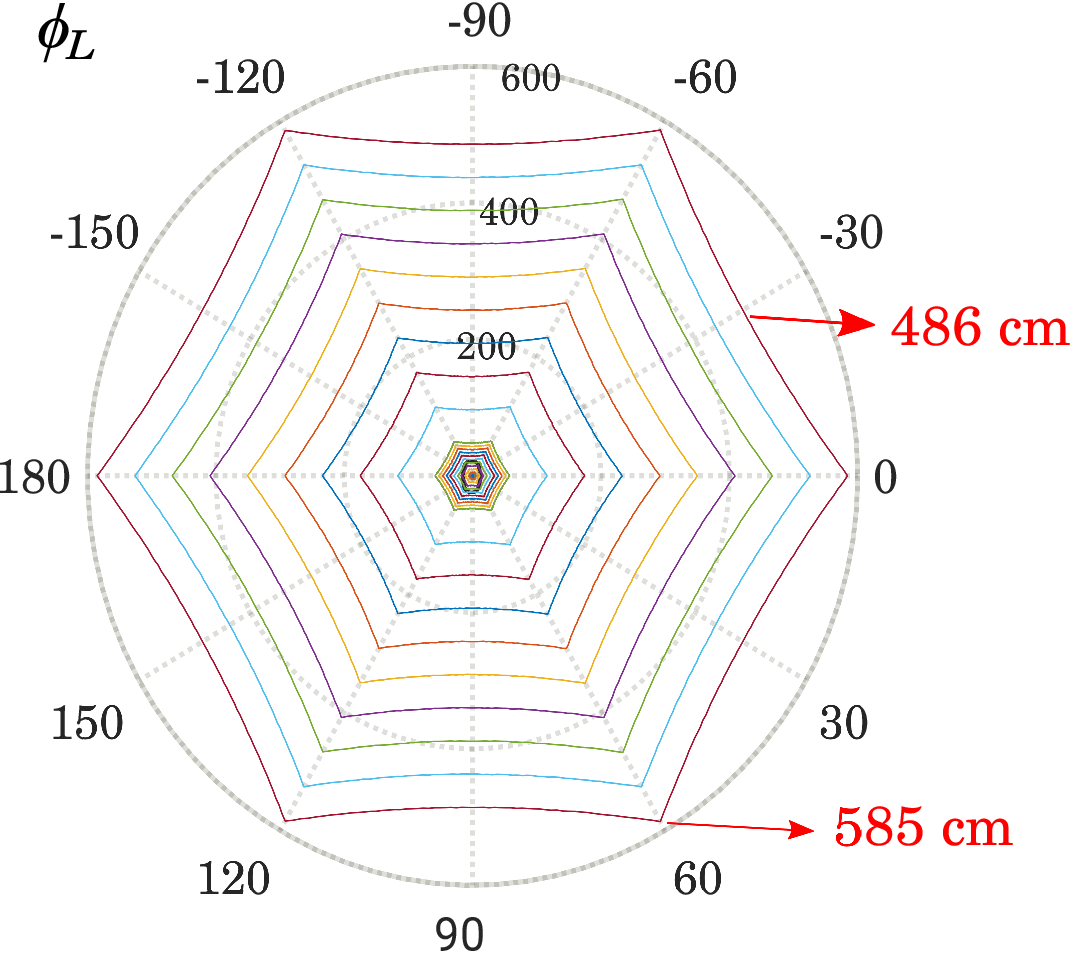}%
\label{fig:figura4b}}
\caption{Drone locations at the limit of the non-ambiguous detector phase range ($\pm$90\textordmasculine) for several drone height ($f$=2.45GHz and $D$=7cm).  Curves traced from 1m to 10m in 1m step and from 1m to 0.1m in 0.1m step. \textbf{(a)}  3D view. \textbf{(b)} Top view where the worst case ($r_{L_{WC}}$=486 cm) and best case ($r_{L_{BC}}$=585 cm) are indicated at 10m drone height.}
\label{fig:figura4}
\end{figure}

Non-ambiguous phase-shift detection range can be exceeded when the drone moves away from the landing point. The higher the drone height ($z_D$), the greater the drone distance ($r_L$). Therefore, for each drone height ($z_D$), maximum separation distance between the drone ($O$) and the landing point is given by non-ambiguous range of phase detector ($\pm$90\textordmasculine). Maximum distances versus drone height ($z_{D_{max}}$=10m) are shown in Fig. \ref{fig:figura4} when $f$=2.45 GHz and $D$=7cm. Roughly, drone tracking can be done in an inverted cone defined by a surface of 486cm radius and the landing point.

The maximum distance (585cm) is about 60\% of the drone height and corresponds to directions given by input points ($P_i$). The maximum distance can be increased by decreasing frequency ($f$) or separation between inputs ($D$). However, if $D$ is increased, the size of the antenna system increases and could make landing maneuvers difficult in the presence of air turbulence. Reducing the frequency also increases the size of the antenna and possibly the minimum necessary separation between them. Furthermore, increasing the maximum distance reduces correction sensitivity expressed in phase detector voltage per cm ($\Delta V_{d}/(2r_{L_{BC}})$. $\Delta V_{d}$=$V_{d_{MAX}}$-$V_{d_{MIN}}$ (Fig. \ref{fig:figura3b}).

\subsection{Landing Algorithm}
\begin{figure}[!t]
\centering
\includegraphics[width=5.8cm]{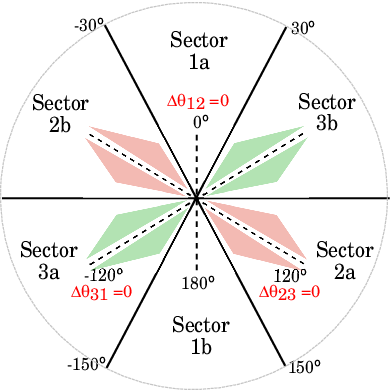}
\caption{Polar coordinates of sectors and corresponding angles obtained from phase-shift curves in Fig. \ref{fig:figura2b}.}
\label{fig:figura5}
\end{figure}

\begin{figure}[!t]
\centering
\includegraphics[width=4cm]{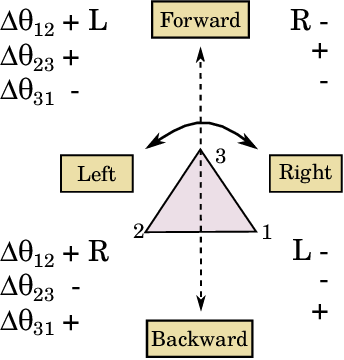}
\caption{Type of multi-rotor maneuvers vs triangular phase-shift sign. The tracking algorithm asumes two type of maneuvers: right/left rotation about the $Z$ axis and forward/backward about $\overline{OP_3}$ direction (Fig. \ref{fig:figura2a}).}
\label{fig:figura6}
\end{figure}

The landing algorithm is based on the phase-shift curves in Fig. \ref{fig:figura2b}. Sectors and angles of interest have been represented in polar coordinates for clarity (Fig. \ref{fig:figura5}). The aim is to provide enough correction instructions to get the drone center ($O$) continuously aligned with the landing point. In this algorithm it has been assumed that drone will perform two types of maneuvers (Fig. \ref{fig:figura6}): right/left rotation about the $Z$ axis and forward/backward about $\overline{OP_3}$ direction (Fig. \ref{fig:figura2a}). 

The maneuver consists of two clearly differentiated parts: rotation to bring the drone to Sector 1, as indicated at the end of section \ref{subseq:Phsvsdrlo}, and maneuvers to track the landing point when the drone is in Sector 1 (right/left rotation and forward/backward movements). This function can be performed as many times as necessary during each descent movement but how the information provided by triangular phase-shift landing circuit applies is not the subject this publication. The algorithm showing the flight maneuvers as a function of the information provided by the circuit for a given height ($z_D$), is detailed in Fig. \ref{fig:figura7}. The flight tracking algorithm to perform the landing maneuver from the phase-shift values is very simple and could be integrated into the same flight control platform, thus avoiding the need to add additional processing elements.

\begin{figure}[!t]
\begin{algorithmic}[1]
\small
\STATE  \textbf{if} \big(\textbf{abs($V_{d_{12}}$)} $\leq$ \textbf{abs($V_{d_{23}}$)}) \& (\textbf{abs($V_{d_{12}}$)} $\leq$ \textbf{abs($V_{d_{31}}$)}\big)
\STATE  \COMMENT{if Sector 1 do nothing}
\STATE  \textbf{else} \COMMENT{Sector 2 or Sector 3}
\STATE  \hspace{0.2cm} \textbf{if} \textbf{abs($V_{d_{23}}$)} $<$ \textbf{abs($V_{d_{31}}$)} 
\STATE  \hspace{0.4cm} 60\textordmasculine Left $\gets$ Yaw Drone \COMMENT{L from Sector 2 to Sector 1}
\STATE  \hspace{0.2cm} \textbf{else} 
\STATE  \hspace{0.4cm} 60\textordmasculine Right $\gets$ Yaw Drone \COMMENT{L from Sector 3 to Sector 1}
\STATE  \hspace{0.2cm} \textbf{end if}
\STATE  \textbf{end if}
\STATE  \COMMENT{L in Sector 1}
\STATE  \COMMENT{Rotate Left/Right \& Move Backward/Forward}    	
\STATE  \textbf{if} \textbf{sign($V_{d_{12}}$)} \textasciitilde= \textbf{sign($V_{d_{23}}$)}
\STATE  \hspace{0.2cm} Right $\gets$ Rotate Drone \COMMENT{L in Right Side}
\STATE  \textbf{else} 
\STATE  \hspace{0.2cm} Left $\gets$ Rotate Drone \COMMENT{L in Left Side}
\STATE  \textbf{end if}
\STATE  \textbf{if} $V_{d_{23}}$ $>$ 0
\STATE  \hspace{0.2cm} Forward $\gets$ Move Drone \COMMENT{L in Sector 1a}
\STATE  \textbf{else}
\STATE  \hspace{0.2cm} Backward $\gets$ Move Drone \COMMENT{L in Sector 1b}
\STATE  \textbf{end if}
\end{algorithmic}
\caption{The flight maneuvers program algorithm as a function of triangular phase-shift tracking information for a given height ($z_D$).}
\label{fig:figura7}
\end{figure}

\section{Triangular Phase-Shift Design}
The phase-shift measurement circuit consists of three phase detectors and their corresponding 90\textordmasculine phase shifters to achieve a non-ambiguous detection range of $\pm$90\textordmasculine\ around the 0\textordmasculine\ reference value (Fig. \ref{fig:figura3}). Furthermore, for the particular case of measuring the response of an array (3 in this case), the phase shifter on each cell would be substituted by a 90\textordmasculine\ hybrid in order to be able to combine the signals of adjacent locations \cite{Pogorzelski2000,Pogorzelski2005}. 

\begin{figure}[!t]
\subfloat[]{\includegraphics[width=4.2cm]{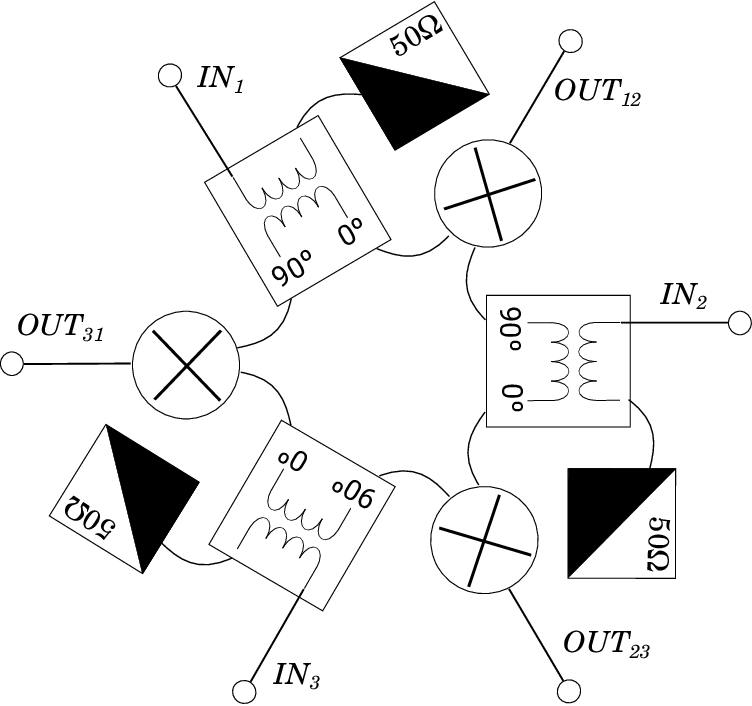}%
\label{fig:figura8a}}
\subfloat[]{\includegraphics[width=4.2cm]{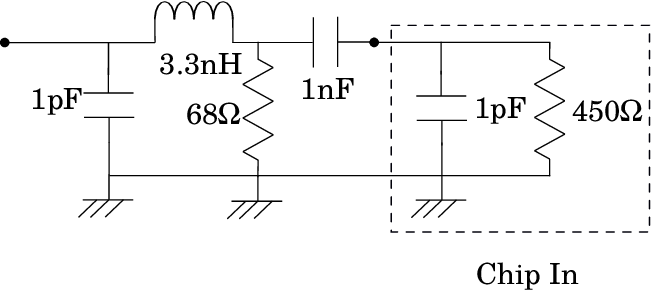}%
\label{fig:figura8b}}
\caption{\textbf{(a)} Simplified schematic of the triangular phase-shift detector for RF drone precise vertical landing. \textbf{(b)} Phase-detector input matching network at 2.45 GHz.}
\label{fig:figura8}
\end{figure}

\begin{figure}[!t]
\centering
\subfloat[]{\includegraphics[width=4.2cm]{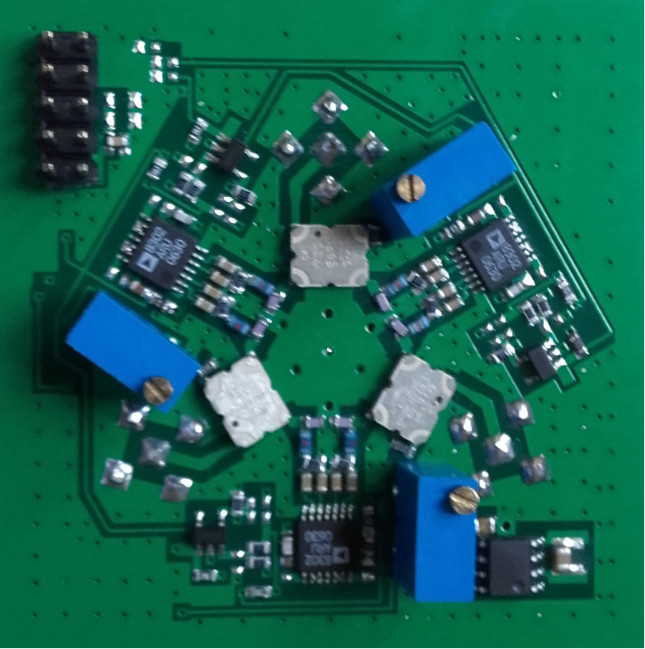}%
\label{fig:figura9a}}
\hfil
\subfloat[]{\includegraphics[width=4.2cm]{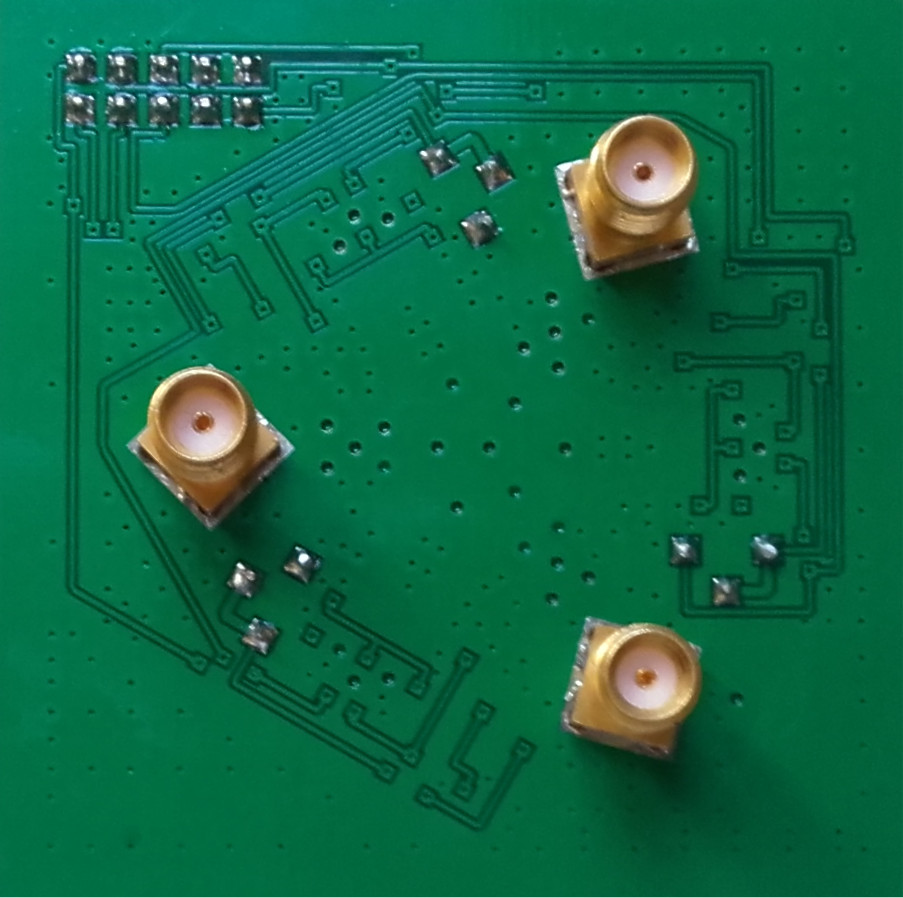}%
\label{fig:figura9b}}
\caption{Prototype of triangular phase-shift detector for RF drone precise vertical landing. \textbf{(a)} Top view. \textbf{(b)} Bottom view.}
\label{fig:figura9}
\end{figure}

Fig. \ref{fig:figura8a} shows a simplified circuit diagram. Triangular arrangement ensures design symmetry and minimizes the effects that can result from mismatching. The input ports ($IN_i$) at every 90\textordmasculine\ hybrid can be moved to distances $D$ (Fig. \ref{fig:figura1}) by wiring. An input matching network has been implemented at 2.45 GHz (Fig. \ref{fig:figura8b}). In addition, an operational amplification stage has been included at the output of each phase detector to adapt and filter the DC voltage, resulting in the final prototype shown in Fig. \ref{fig:figura9}. Commercial devices have been used in the design prototype: analog phase detectors (AD8302) with LF (Low Frequency) to 2.7GHz frequency band, high input dynamic range (-60 dBm to 0 dBm) and slope of 10 mV/\textordmasculine; hybrid couplers (X3C26P1-03S) with a phase-shift of 90\textordmasculine$\pm$4\textordmasculine\ at 2.3 GHz to 2.9 GHz.  Equation \ref{eq:phd_vd} shows the theoretical triangular characteristic of the phase-detector output voltage AD8302. 
\begin{eqnarray}
\label{eq:phd_vd}
V_{d_{ij}} & = &  K_d ( 180 - |\Delta \theta_{ij}|)\\
K_d        & = & 10 (\textrm{mV/\textordmasculine }) \nonumber
\end{eqnarray}

\section{Characterization and Validation}

The purpose of the circuit characterization is to obtain the expressions that allow converting output voltages in each detector to the phase-shifts that can be processed by the landing algorithm detailed in Fig. \ref{fig:figura7}. The characterization of the manufactured prototype is composed of three different parts. The first step is to calibrate the measuring system to ensure that both generators are well phase-synchronized and that the reference planes are located on cable connectors. In the second step, measurements are made by varying amplitude and phase-shift of the input signals to the prototype, in order to obtain both the phase-shift as function of the voltages measured for each detector output, and the non-ambiguity ranges to determine the maximum drone distance deviation to landing point. Finally, the flight maneuvers algorithm is validated through values obtained from circuit characterization. 

\subsection{Calibration of the measurement system}

\begin{figure}[!t]
\centering
\includegraphics[width=8.8cm]{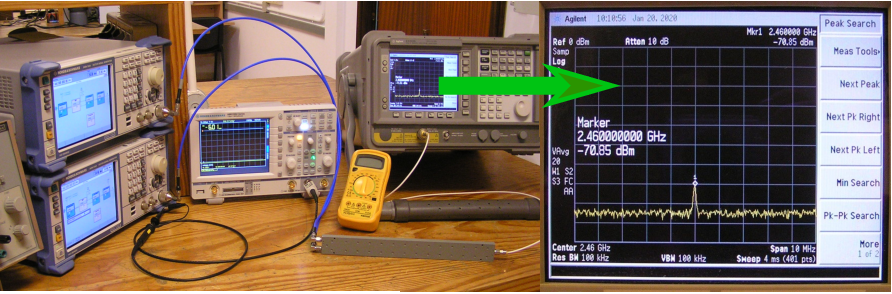}
\caption{Connections of signal vector generators (R\&S SMB-2345-A) for measurement system calibration.}
\label{fig:figura10}
\end{figure}

The calibration of the measurement system must ensure that the amplitudes and phase-shifts of the input signals correspond to those under flight conditions. The detailed procedure is described in \cite{Umpierrez2012} but a few simplifications have been made. Because the distances traveled by RF signal between the landing point and the input points ($P_i$) are very similar, the generators are only calibrated for equal amplitudes. In summary, back connections between generators are made to share both reference and conversion oscillators. The phase of the master generator is set to 180\textordmasculine\ and its output amplitude to a specific value. Amplitude calibration is performed by means of a power meter located at the output connector of the cable that is connected to the master generator. Then, both generators are connected to the inputs of a power combiner and its output is connected to a spectrum analyzer. The phase and amplitude of the slave generator is modified until the power combiner output is minimized. This process must be performed for each power and frequency. Fig. \ref{fig:figura10} shows equipment, connections and the minimum measured value (-70.85 dBm) when frequency is 2.46 GHz and power is -20 dBm. At this point, both amplitudes and phase-shifts between generators are well known in the reference plane located at the end of cable connectors.

\subsection{Triangular phase-shift circuit}

\begin{figure}[!t]
\centering
\subfloat[]{\includegraphics[width=8cm]{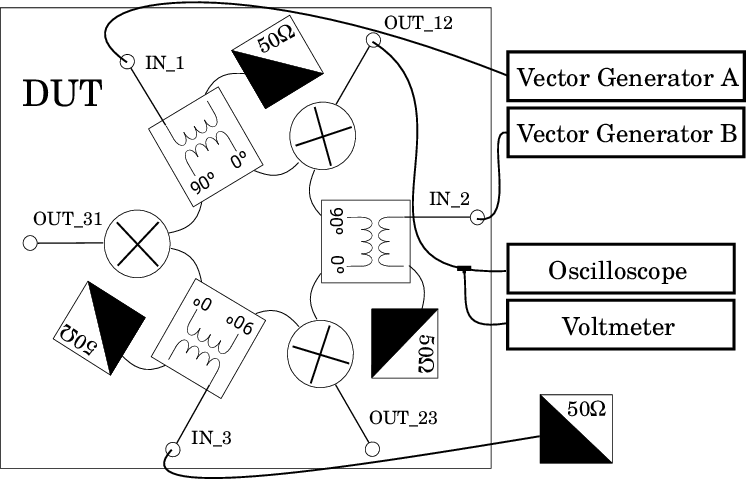}
\label{fig:figura11a}}
\hfil
\subfloat[]{\includegraphics[width=8cm]{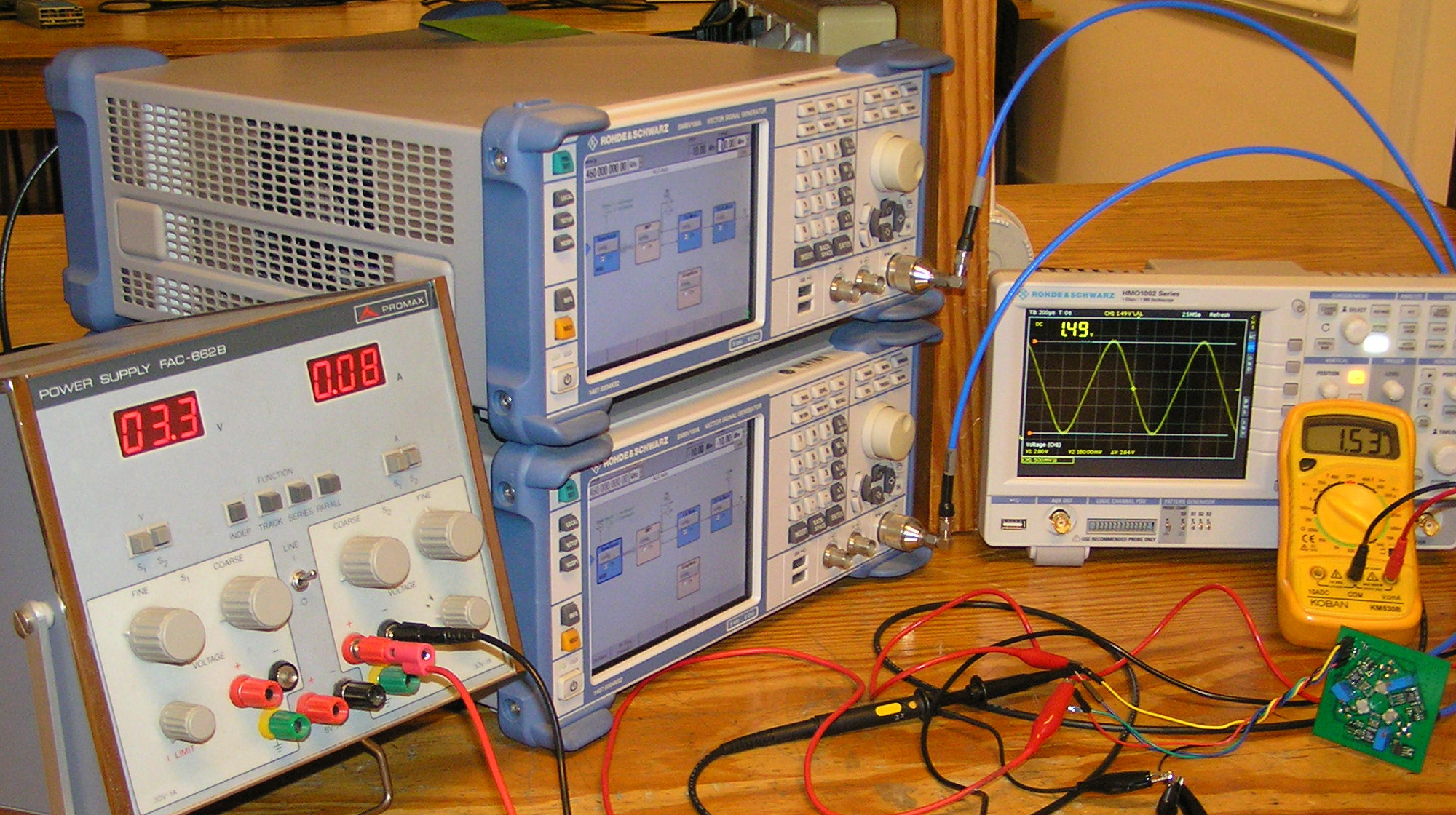}
\label{fig:figura11b}}
\caption{Measurement system for the characterization of the prototype of triangular phase-shift sensor ($f$=2.46 GHz). \textbf{(a)} Schematic,
\textbf{(b)} Laboratory Assembly}
\label{fig:figura11}
\end{figure}

The calibrated signals are applied to the circuit and potentiometers are adjusted to maximize the detector output voltage range when the prototype is connected to 3.3V power supply. Measurements are made two by two, loading the unconnected input with 50$\Omega$. To obtain the curve of the detector, a 1kHz offset is added to one of the generators. The complete assembly of the measuring system is shown in Fig. \ref{fig:figura11}. Moreover, the detector signal aspect of the phase-shift detector is visualized in the oscilloscope screen when the input amplitude is -20dBm. Under these conditions, the detector output voltage measured on the oscilloscope varies from 140mV to 2.84V (central value = 1.49V). Finally, the circuit characterization is done at 2.46 GHz, where the mismatching of the coupling network makes the central value correspond to zero value in $\Delta \theta_{12}$ phase-shift.

Then, the 1kHz offset between the synchronized generators is removed and the phase of one of them is varied to measure the output voltages of each detector using a voltmeter. The curves of each detector show great similarity when the input amplitude varies between -10 dBm and -40 dBm. Table \ref{tab:table_i} shows measurements when the input amplitude is -20 dBm. The $V_{d_{REF}}$ is the detector voltage when the input phase-shitf is zero. Measurements show a displacement of detector curves mainly due to variation in frequency response of the used components and mismatching \cite{Perez2017}. The largest deviation occurs in the output corresponding to the  phase-shift $\Delta \theta_{31}$ (+7\textordmasculine). If a symmetrical range is assumed, the maximum phase-shift that can be considered is $\pm$83\textordmasculine. However, a non-ambiguous range value of $\pm$80\textordmasculine\ is used in the prototype characterization to avoid range limits and prevent possible deviations.

\begin{figure}[!t]
\centering
\includegraphics[width=7.5cm]{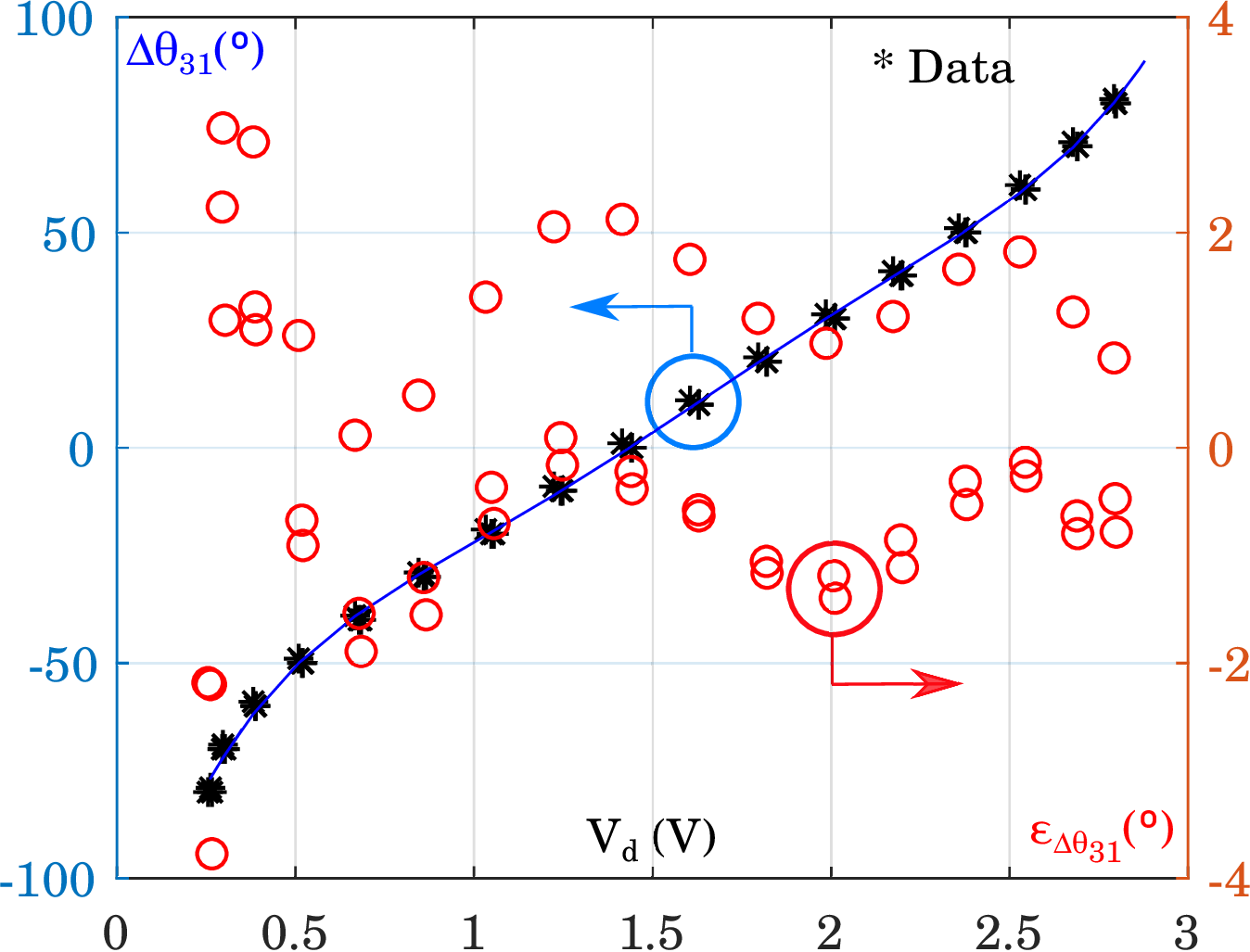}
\caption{Representation of the fifth degree polynomial function that has been used to model the response of the detector that measures the phase-shift $\Delta \theta_{31}$. In addition, the measured data and phase-shift error are included. The polynomial is valid in the non-ambiguous phase shift range of $\pm$80\textordmasculine.}
\label{fig:figura12}
\end{figure}

\begin{figure}[!t]
\subfloat[]{\includegraphics[width=3.4cm]{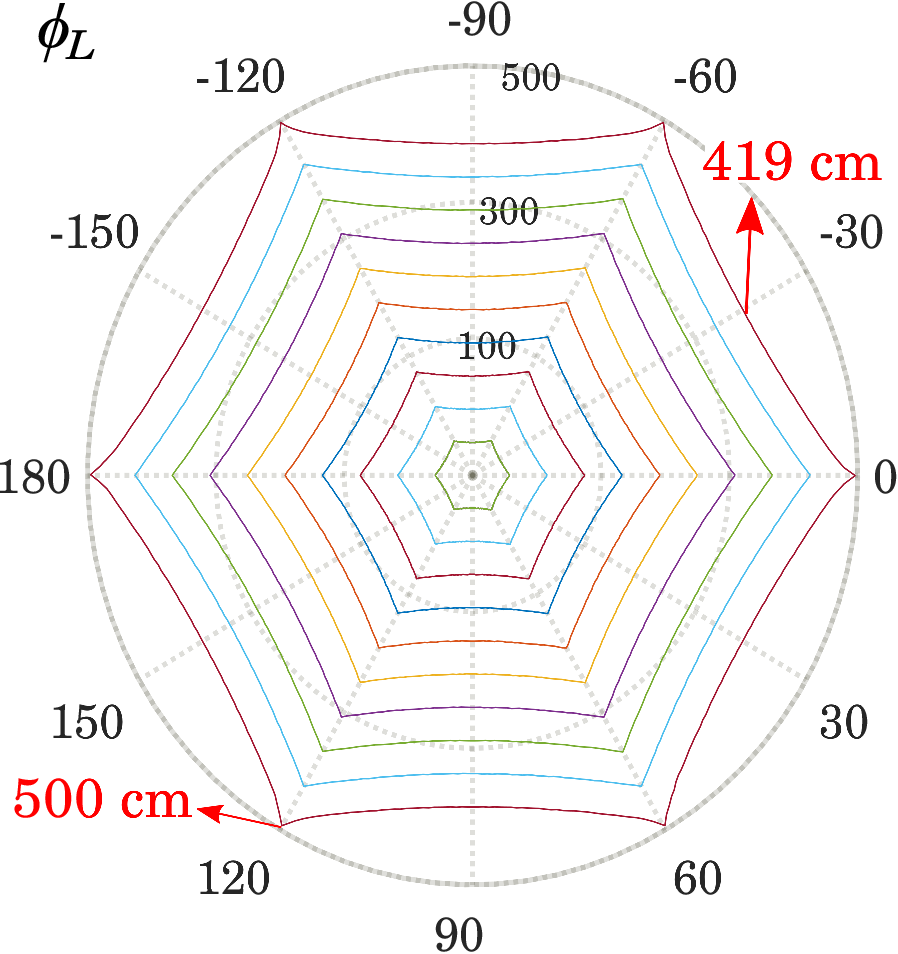}
\label{fig:figura13a}}
\subfloat[]{\includegraphics[width=5.3cm]{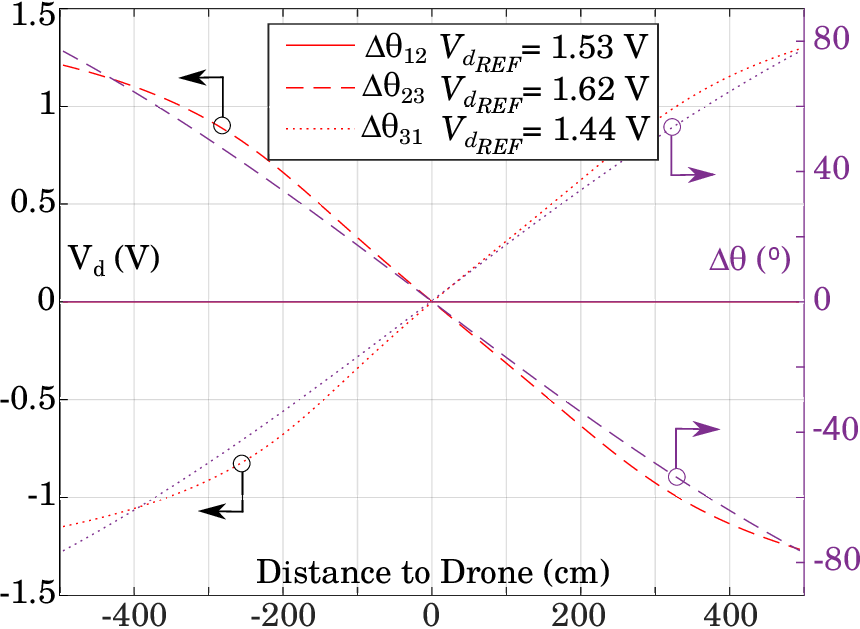}%
\label{fig:figura13b}}
\caption{Prototype characterization. \textbf{(a)} Drone locations at the limit of the non-ambiguous detector phase range ($\pm$80\textordmasculine) for several drone height ($f$=2.46GHz and $D$=7cm). Curves traced from 1m to 10m in 1m step: worst case ($r_{L_{WC}}$=419 cm) and best case ($r_{L_{BC}}$=500 cm) for 10m are highlighted. \textbf{(b)}  Phase-shifts ($\Delta \theta_{ij}$ and detector voltages ($V_{d_{ij}}$) when drone moves along the $Y$ axis and landing point is located in \{0,0,0\}.}
\label{fig:figura13}
\end{figure}

An n$^{th}$ degree polynomial is chosen to model the detector response because measured curve does not correspond to the device theoretical one (Eq. \ref{eq:phd_vd}) nor to the multiplier analog sinusoidal one. As an example, Table \ref{tab:table_ii} shows the fifth degree polynomial coefficients values and the maximum error obtained in each detector when the input amplitude varies between -10dBm and -40dBm. The maximum error is given by the detector that defines the non-ambiguous range, that is, the one corresponding to $\Delta \theta_{31}$ (3.8\textordmasculine). Fig. \ref{fig:figura12} shows the Phase-shift curve ($\Delta \theta_{13}$) vs $V_{d}$, the data and the error at each measurement point when the input signal varies between -10dBm and -40dBm. The error is higher in the curved line of the phase-shift detectors (phase-shifts close to -80\textordmasculine) and mainly in $V_{d_{23}}$ and $V_{d_{31}}$ (Table \ref{tab:table_ii}). Fig. \ref{fig:figura13a} shows the reduction of the drone-tracking inverted cone volume compared to the one generated in Fig. \ref{fig:figura4}. This occurs mainly due to the changes in the non-ambiguous range (from $\pm$90\textordmasculine to $\pm$80\textordmasculine) although there is also a slight frequency effect (from $f$=2.45GHz to $f$=2.46GHz). The maximum separation decreases from about 60\% (585cm @2.45GHz) to 50\% (500cm @2.46GHz) of the drone height (10m) in the direction of the axes fixed by the input points ($P_i$). In order to make an assessment of the interpolation error, Fig. \ref{fig:figura13b} shows the phase-shift and voltage detected as a function of the distance to the landing point for a worst case trajectory: $L$ located at the point \{0,0,0\} when the drone moves along the $Y$ axis. In that case, the phase-shift between point $P_1$ and $P_2$ is zero ($\Delta \theta_{12}$=0) and the other two vary from -80\textordmasculine\ to +80\textordmasculine. According to Fig. \ref{fig:figura13b} ($\phi_L$=-180\textordmasculine\ and $\phi_L$=0\textordmasculine), the drone can move through the non-ambiguous zone from -5m to 5m when the drone is at the height of 10m. The curves show a linear relationship between phase-shift and distance. For -80\textordmasculine, the error shown in Fig. \ref{fig:figura12} is almost 4\textordmasculine, which means an uncertainty of 2.5\% (4/[2$\cdot$80]$\cdot$100) in the range $\pm$80\textordmasculine, that is, $\pm$25 cm over 10m ($\pm$5m) in the worst case. However, the error is in most cases less than 2\textordmasculine, that is, less than 1.25\%. These percentages are maintained for other heights since the non-ambiguous zone ($\pm$80\textordmasculine) does not vary. 

\begin{table}[!t]
\caption{Values measured at detector outputs when the input power is -20 dBm and $f$=2.46GHz}
\label{tab:table_i}
\centering
\begin{tabular}{|c|c|c|c|c|c|c|}
\hline
3.3V 80mA & 
\multicolumn{2}{c|}{$\mathbf{\Delta \theta_{12}}$} & \multicolumn{2}{c|}{$\mathbf{\Delta \theta_{23}}$} &
\multicolumn{2}{c|}{$\mathbf{\Delta \theta_{31}}$}\\ \hline
 & -100\textordmasculine & 0.228V & -100\textordmasculine & 0.253V & -100\textordmasculine & 0.352V\\ \hline
 & -90\textordmasculine & 0.197V & -90\textordmasculine & 0.248V & -90\textordmasculine & 0.283V\\ \hline
 & -80\textordmasculine & 0.223V & -80\textordmasculine & 0.299V & -80\textordmasculine & 0.266V\\ \hline
 & -70\textordmasculine & 0.302V & -70\textordmasculine & 0.399V & -70\textordmasculine & 0.303V\\ \hline
 \rowcolor{cyan}
 $V_{d_{REF}}$ &   0\textordmasculine & 1.533V &   0\textordmasculine & 1.610V &  0\textordmasculine & 1.443V\\ \hline
 &   0\textordmasculine & 1.533V & -4\textordmasculine & 1.571V & +7\textordmasculine & 1.577V\\ \hline
 & 70\textordmasculine & 2.756V & 70\textordmasculine & 2.814V & 70\textordmasculine & 2.691V\\ \hline
 & 80\textordmasculine & 2.837V & 80\textordmasculine & 2.873V & 80\textordmasculine & 2.796V\\ \hline
 & 90\textordmasculine & 2.865V & 90\textordmasculine & 2.879V & 90\textordmasculine & 2.854V\\ \hline
 & 100\textordmasculine & 2.838V & 100\textordmasculine & 2.832V & 100\textordmasculine & 2.863V\\
\hline
\end{tabular}
\end{table}

\begin{table}[!t]
\caption{Interpolation polynomial coefficients at 2.46 GHz when the input power varies from -10 to -40 dBm and
$\Delta \theta$ = $F(V_{d})$ in $\pm$80\textordmasculine. The error is defined as $\epsilon_{\Delta \theta}$ =
$F (V_{d})$ - Data.}
\label{tab:table_ii}
\centering
\begin{tabular}{|c|c|c|c|}
\hline
Coeff. & $\mathbf{\Delta \theta_{12}}$ & $\mathbf{\Delta \theta_{23}}$ & $\mathbf{\Delta \theta_{31}}$\\ \hline
$a_0$ & -114.203 &  -125.812 &  -129.954 \\ \hline 
$a_1$ & 199.396 & 211.489 & 274.718 \\ \hline
$a_2$ & -228.453 &  -240.403 &  -328.593 \\ \hline 
$a_3$ & 164.691 & 172.357 & 226.222\\ \hline
$a_4$ & -55.965 & -58.608 & -73.488 \\ \hline   
$a_5$ & 7.245 & 7.596 & 9.115\\ \hline
$|\epsilon_{\Delta \theta}|$ & $\leq$ 0.9\textordmasculine & 
$\leq$ 1.7\textordmasculine & $\leq$ 3.8\textordmasculine\\ \hline
$V_{d_{REF}}$ & 1.530V & 1.624V & 1.436V\\ \hline 
\end{tabular}
\end{table}

\subsection{Flight maneuvers algorithm}
The flight maneuvers algorithm detailed in Fig. \ref{fig:figura7} has been validated by several examples where a detailed analysis of the maneuvers is derived from the values of the voltages provided by the Table \ref{tab:table_ii} interpolation polynomials. The landing point is placed in an arbitrary position relative to the drone within the non-ambiguous tracking zone, the phase-shifts are calculated from equations \ref{eq:deltaphi} to \ref{eq:deltas} and the detected voltages are obtained from the Table \ref{tab:table_i} interpolation polynomials. The voltage values are used in the algorithm to determine the maneuver to be performed. This process is done continuously until the drone is located just above the landing point. It has been assumed that this situation occurs when $|V_{d_{ij}}| <$ 0.02V. In addition, the value of 1cm has been used for the forward/backward maneuvers (MD=1cm) and 1\textordmasculine\ for the right/left turning maneuver (RD=1\textordmasculine). 

\begin{figure}[!t]
\centering
\includegraphics[width=8.8cm]{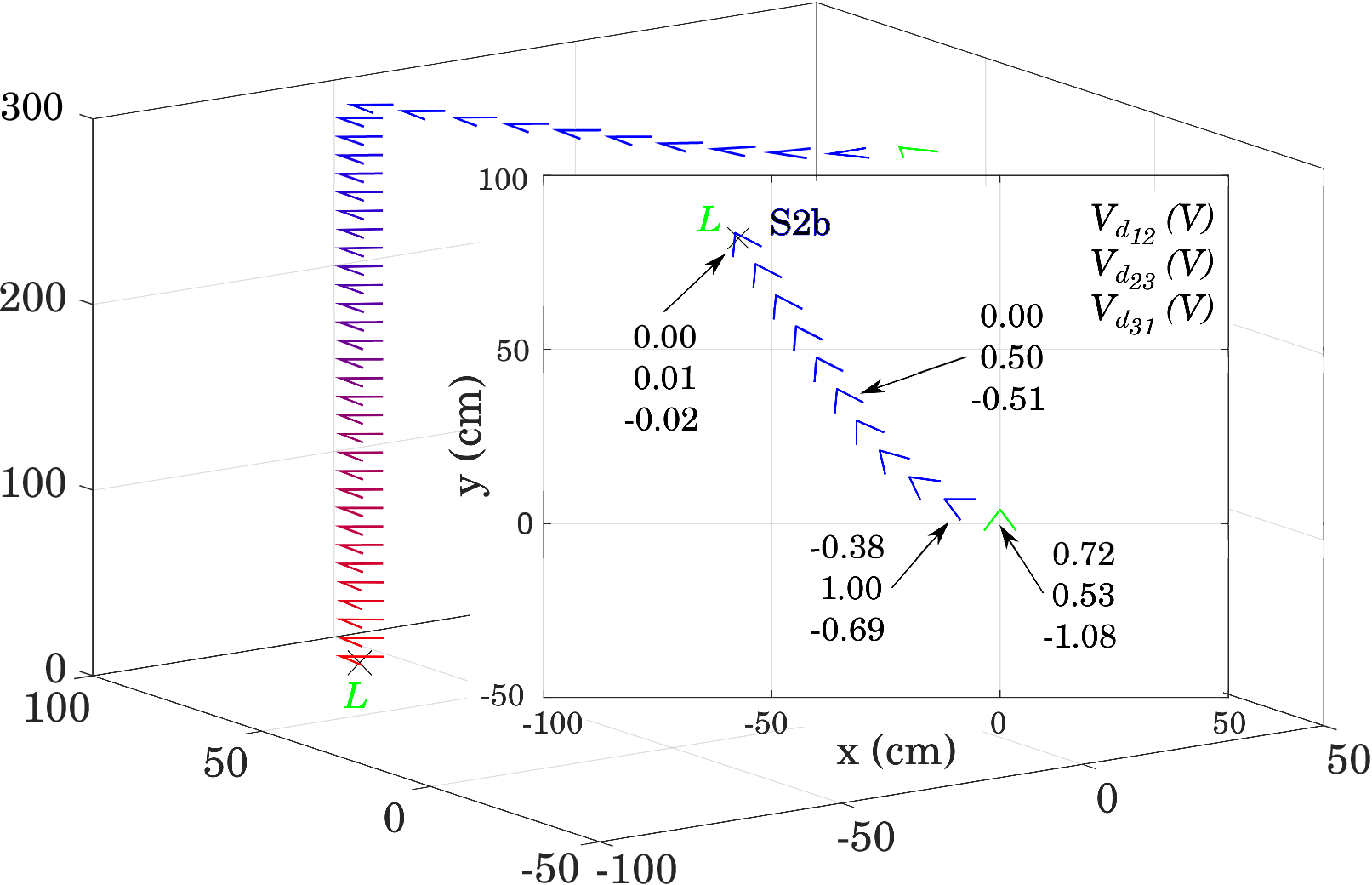}
\caption{Drone trajectory when the landing point ($L$) is located in Sector 2b. ($\phi_L$=-35\textordmasculine, $r_L$=100cm). Drone located in \{0,0,300\}. $XY$ view is added for clarity. MD=1cm and RD=1\textordmasculine\ (one in ten values is displayed).}
\label{fig:figura14}
\end{figure}

In Fig. \ref{fig:figura14}, the drone (in green color) has been located at the point \{0,0\}, oriented according to the positive $Y$ axis (Fig. \ref{fig:figura2a}) and height $z_D$=300cm. The landing point has been located in Sector 2b ($\phi_L$=-35\textordmasculine\ and $r_L$=100cm). The detector voltages provided by the interpolation polynomial ($V_{d_{12}}$=0.72V, $V_{d_{23}}$=0.53V, $V_{d_{31}}$=-1.08V) correspond to $L$ point location in Sector 2b. According to the algorithm, a 60\textordmasculine\ turn is made to the left, which guarantees that point $L$ is in Sector 1. The next values ($V_{d_{12}}$=-0.38V, $V_{d_{23}}$=1.00V, $V_{d_{31}}$=0.69V) involve a combined forward and right movement. Two additional situations are detailed: an intermediate point corresponding to forward and left movement in Sector 1a ($V_{d_{12}}$=0.00V, $V_{d_{23}}$=0.50V, $V_{d_{31}}$=-0.51V), and location prior to descent when the drone has been placed just over the point $L$ ($V_{d_{12}}$=0.00V, $V_{d_{23}}$=0.01V, $V_{d_{31}}$=-0.02V). The complete landing trajectory has been represented considering a descent of 1cm in each maneuver. Although Fig. does not illustrate corrections in detail, the algorithm is continuously proposing right/left and forward/backward maneuvers.

\begin{figure}[!t]
\centering
\includegraphics[width=8.8cm]{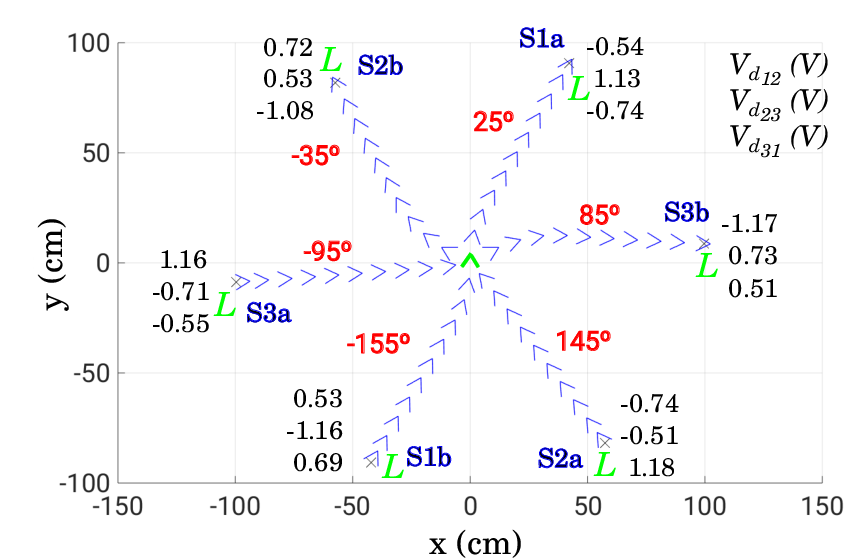}
\caption{Drone trajectory when the landing point ($L$) is located in near the edge of each sector. Landing point locations and initial detector voltages are detailed for each case. MD=1cm and RD=1\textordmasculine\ (one in ten values is displayed).}
\label{fig:figura15}
\end{figure}

Fig. \ref{fig:figura15} shows several trajectories with point $L$ in different sectors. It includes the detector voltages when the drone is at the starting point (\{0,0,300\} in green) corresponding to the different positions of the landing point. In all cases and following the sequence of the algorithm in Fig. \ref{fig:figura7}, the drone is guided until it is just above point $L$, being able to start the descent from that moment, executing the algorithm for each height.

\section{Conclusions}
The triangular phase-shift detector has been designed to facilitate landing maneuvers when precise pointing is required over the landing point. It is a circuit especially indicated in the last landing meters since it can provide a large amount of data with a very low computational cost. The design is particularly suitable for the final metres of abrupt, forested and windy areas where it is necessary to have a high data rate that allows corrections to be made at high speed. Combined with imaging and GPS it can provide greater robustness to accurate vertical landing systems. However, directivity of the antennas, RF power of the oscillator at the landing point, or sensitivity of the triangular phase-shift circuit will limit the final maximum height. The system only needs a frequency oscillator placed at the landing point and a reception platform (antennas). The data rate used to position the drone with respect to the landing point would be determined by the processing time of voltages provided by the circuit.
The circuit performance is analyzed according to the design parameters: frequency, separation between the receiver elements and drone height, among others. An algorithm is displayed that uses the phase-shift information provided by the circuit and allows the drone to perform maneuvers aimed to align the drone with the landing point. Simplicity of the algorithm and the polynomial expression used to compute phase-shifts, suggest the possibility of integration in the drone flight controller, with no need of additional processing systems.

To demonstrate the feasibility of the proposed design, a triangular phase-shift detector prototype has been implemented using commercial devices: analog detectors with high input dynamic range at LF to 2.7 GHz and hybrid couplers with a phase-shift of 90\textordmasculine$\pm$4\textordmasculine\ at 2.3 GHz to 2.9 GHz. Once adjusted, it is characterized at 2.46 GHz where the circuit shows a better performance. The prototype has a non-ambiguous phase-shift detection range of $\pm$80\textordmasculine, allowing the drone to be tracked on about circular surface with a radius equivalent to 50\% of the drone height. A fifth degree polynomial function allows to compute the phase-shifts as a function of the output voltages of each detector with an error less than 3.8\textordmasculine\ in the worst case, when the RF input signal varies from -10dBm to -40dBm. 


%



\section*{Acknowledgment}
Thanks to Juan Domingo Santana Urb\'in for his enormous contribution during this research.

\ifCLASSOPTIONcaptionsoff
  \newpage
\fi



\bibliographystyle{IEEEtran}
\bibliography{mybibliography.bib}

%
\vspace{-1cm}
\begin{IEEEbiography}[{\includegraphics[width=1in,height=1.25in,clip,keepaspectratio]{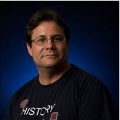}}]{V\'ictor Ara\~na-Pulido} (A'94-M'05) was born in Las Palmas de Gran Canaria, Spain, in 1965. He received the Telecommunication Engineering degree from the Universidad Polit\'ecnica de Madrid, Madrid, Spain in 1990, and the Ph.D. degree from the Universidad de Las Palmas de Gran Canaria (ULPGC), Las Palmas, Spain, in 2004. He is currently an Assistant Professor with the Signal and Communication Department and member of IDeTIC-ULPGC. He has been the leading Researcher in several Spanish research and development projects and has taken part in a number of Spanish and European projects in collaboration with industries. His research interests include the nonlinear design of microwave circuits, control subsystem units, and communications systems applied to data acquisition complex networks.
\end{IEEEbiography}
\vspace{-1cm}
\begin{IEEEbiography}[{\includegraphics[width=1in,height=1.25in,clip,keepaspectratio]{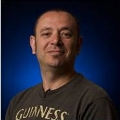}}]{Eugenio Jim\'enez-Ygu\'acel}(A'94-M'05). He received an M.Sc. and a Ph.D. from the Universidad Polit\'ecnica de Madrid, Madrid, Spain, in 1991 and 2002, 
respectively. He joined the University of Las Palmas de Gran Canaria, Las Palmas, Spain, in 1993, where he is currently an Assoc. Professor and member of IDeTIC-ULPGC. His initial activity in R\&D focuses on the design, manufacture and measurement of antennas and RF and microwave circuits. Along these lines, he has participated in national public projects uninterruptedly since 1988 when, as a student, he joined the Radiation Group of the UPM. In 2006 he began a new line related to digital communications in HF (1-30 Mhz) and signal processing through programmable logic. That line continues today trying to open new fronts such as submarine communication using radio frequency.
\end{IEEEbiography}
\vspace{-1cm}
\begin{IEEEbiography}[{\includegraphics[width=1in,height=1.25in,clip,keepaspectratio]{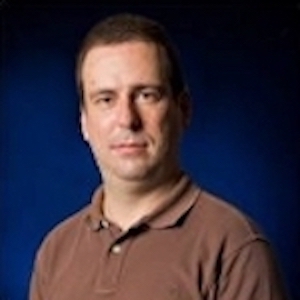}}]{Francisco Cabrera-Almeida} (A'98-M'05) was born in Las Palmas de Gran Canaria, Spain, in 1970. He received his Telecommunications Engineering degree in 1997 from the Universidad de Las Palmas de Gran Canaria (ULPGC), Spain, and his Ph.D. degree from ULPGC, in 2012. He is currently an Assistant Professor with the Signal and Communications Department and member of the Institute for Technological Development and Innovation in Communications (IDeTIC-ULPGC) since it was founded in 2010.He has taken part in a number of Spanish and European projects in collaboration with industries and other universities. His research interests include: numerical electromagnetic modeling techniques, radiowave propagation and communications systems applied to data acquisition complex networks. 
\end{IEEEbiography}
\vspace{-1cm}
\begin{IEEEbiography}[{\includegraphics[width=1in,height=1.25in,clip,keepaspectratio]{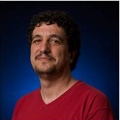}}]{Pedro Quintana-Morales}
was born in Santa Cruz de Tenerife, Spain, in 1964. He received the Telecommunication Engineering degree from the Universidad Polit\'ecnica de Madrid, Madrid, Spain in 1989, and the Ph.D. degree from the Universidad de Las Palmas de Gran Canaria (ULPGC), Las Palmas, Spain, in 2016. He is currently an Assistant Professor with the Signal and Communication Department and member of IDeTIC-ULPGC. He has taken part in a number of Spanish and European projects in collaboration with industries and other universities. His research interests include signal processing, stochastic processes and data analysis applied to speech, images, biosignals, radio communications and sensor networks.
\end{IEEEbiography}




\end{document}